\documentclass[12pt]{iopart}
\usepackage{psfig}% Include figure files
\usepackage{graphicx}
\usepackage{graphics}
\begin{document}

\title{Transverse Pressure and Strangeness Dynamics in Relativistic Heavy Ion Reactions}

\author{M. Bleicher, E. Bratkovskaya, S. Vogel, X. Zhu}

\address{Institut f\"ur Theoretische Physik,
J.~W.~Goethe Universit\"at, 60054 Frankfurt am Main,
Germany}
\address {FIAS, Frankfurt Institute for Advanced Studies, Robert-Mayer-Str. 10, 60054 Frankfurt am Main, Germany}

\begin{abstract}
Transverse hadron spectra
from proton-proton, proton-nucleus and nucleus-nucleus collisions from
2 AGeV to 21.3 ATeV are investigated within two independent transport
approaches (HSD and UrQMD). For
central Au+Au (Pb+Pb) collisions at energies above $E_{\rm lab}\sim$ 5
AGeV, the measured $K^{\pm}$ transverse mass
spectra have a larger inverse slope parameter than expected from the
default calculations. The additional pressure
- as suggested by lattice QCD calculations at finite quark chemical
potential $\mu_q$ and temperature $T$ - might be generated by strong
interactions in the early pre-hadronic/partonic phase of central Au+Au
(Pb+Pb) collisions. This is supported by a non-monotonic energy dependence 
of $v_2/\langle p_T\rangle$ in the present transport model.
\end{abstract}

\maketitle

\section*{Introduction}

Recent lattice QCD (lQCD) calculations at vanishing quark chemical potential and
finite temperature indicate critical energy densities for the formation
of a quark-gluon plasma (QGP) of $\sim$ 0.6-1 GeV/fm$^3$ \cite{Karsch}.
Such energy densities might already be achieved at Alternating Gradient Synchrotron
(AGS) energies of $\sim$ 10 AGeV for central Au+Au collisions
\cite{HORST,exita,Weber98}. According to lQCD calculations at finite 
quark chemical potential $\mu_q$ \cite{Fodor}
a rapid increase of the thermodynamic pressure $P$ with
temperature above the critical temperature $T_c$ for a cross over
(or phase transition) to the QGP is expected.

\begin{figure}[hbt]
\vspace*{.2cm}
\centerline{\psfig{figure=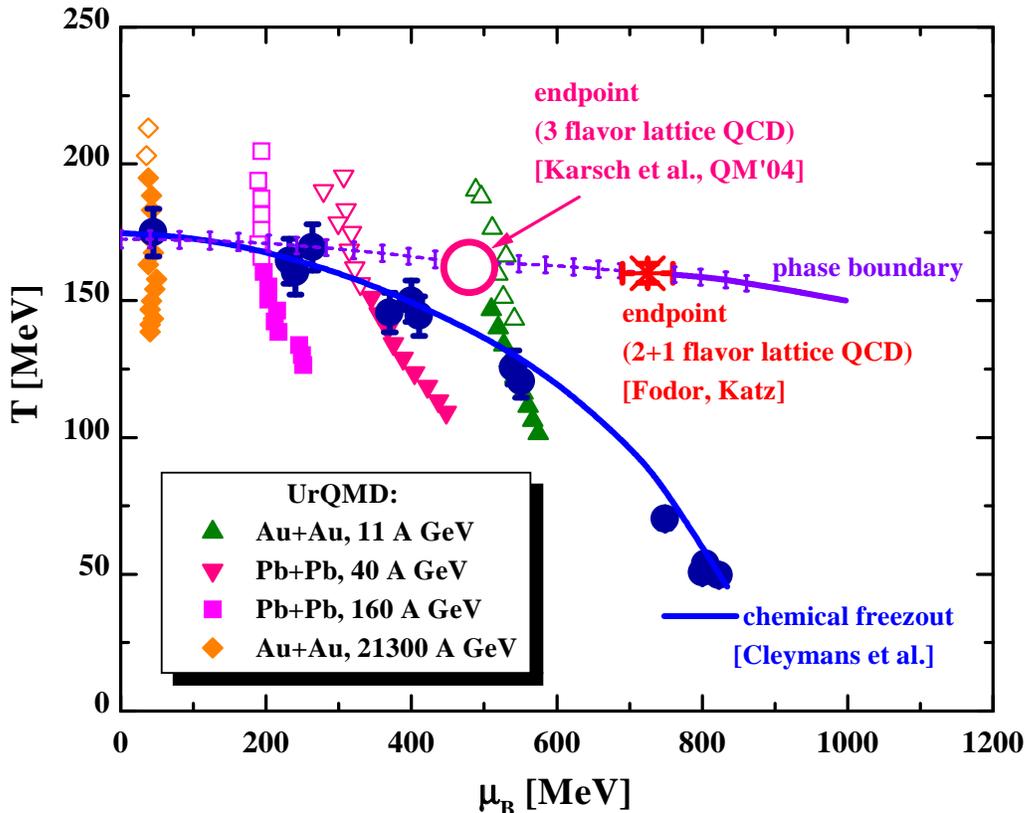,width=13.5cm}}
%\insertplot{tmuurqmd.eps}
\vspace*{-.5cm}
\caption{ Schematic phase diagram in the $T-\mu_B$ plane.  The solid
line characterises the universal chemical freeze-out line from Cleymans
et al. \protect\cite{Cleymans} and the full dots (with error bars)
denote the 'experimental' chemical freeze-out parameters from Ref.
\protect\cite{Cleymans}. The various symbols represent temperatures $T$
and chemical potentials $\mu_B$ extracted from UrQMD 1.3 transport
calculations in central Au+Au (Pb+Pb) collisions at 21.3 A$\cdot$TeV,
160, 40 and 11 A$\cdot$GeV \protect\cite{Bravina} (see text). The large
open circle and the star indicate the tri-critical endpoints and phase boundary from
lattice QCD calculations by Karsch et al.  \protect\cite{Karsch2} and
Fodor and Katz \protect\cite{Fodor}, respectively. The horizontal line
with error bars is the phase boundary from \protect\cite{Fodor}.}
\label{Fig_QCD}
\end{figure}

Following the previous study \cite{MT-prl} we speculate that
partonic degrees of freedom might be responsible for this effect
already at $\sim$ 5 $A\cdot$GeV. Our arguments here are based on a
comparison of the thermodynamic parameters $T$ and $\mu_B$ extracted
from the transport models in the central overlap regime of Au+Au
collisions \cite{Bravina} with the experimental systematics on chemical
freeze-out configurations \cite{Cleymans} in the $T,\mu_B$ plane. The
solid line in Fig.  \ref{Fig_QCD} characterises the universal chemical
freeze-out line from Cleymans et al. \cite{Cleymans} and the full
dots with error bars denote the 'experimental' chemical freeze-out
parameters - determined from the thermal model fits to the experimental
particle ratios \cite{Cleymans}. The various smaller symbols (in vertical
sequence) represent temperatures $T$ and chemical potentials $\mu_B$
extracted from UrQMD 1.3 transport calculations in central Au+Au
(Pb+Pb) collisions at $\sqrt s=200$~AGeV, $E_{\rm lab}= 160$, 40 and 11 A$\cdot$GeV
\cite{Bravina} as a function of the reaction time in the center-of-mass
(from top to bottom).

During the non-equilibrium phase (open symbols) the transport
calculations show much higher temperatures (or energy densities) than
the 'experimental' chemical freeze-out configurations at all bombarding
energies ($\geq$~11~A$\cdot$GeV).  These numbers are also higher than
the tri-critical endpoints and phase boundary extracted from lattice QCD calculations by
Karsch et al.  \cite{Karsch2} (large open circle) and Fodor and Katz
\cite{Fodor} (star with horizontal error bar). Though the QCD lattice
calculations differ substantially in the value of $\mu_B$ for the
critical endpoint, the critical temperature $T_c$ is close to
160 MeV in both calculations, while the energy density is of the order
of 1 GeV/fm$^3$ or even below. This diagram shows that at
RHIC energies one encounters more likely a cross-over between the
different phases when stepping down in temperature during the expansion
phase of the hot fireball. 

Indeed, a {\it hardening}  of the 
measured transverse mass ($m_t$) spectra in
central Au+Au collisions relative to pp interactions \cite{NA49_T,Goren} 
from  AGS energies on is observed.
This increase of the inverse slope parameter $T$ is commonly
attributed to strong collective flow, which is absent in the respective
pp or pA collisions.
It has been proposed \cite{SMES} to interpret the
high and approximately constant $K^\pm$ slopes above $\sim 30$ AGeV --
the 'step' -- as an indication of the phase transition. 

In this contribution we  explore whether the pressure needed to
generate a large collective flow to explain the hard slopes of the
$K^\pm$ spectra with a 'plateau' at SPS energies is produced in the 
present transport models by interactions of hadrons or whether additional 
partonic contributions in the equation of state might be needed to explain these 
effects (for further details the reader is referred to \cite{Bratkovskaya:2004kv}).
To understand whether a failure of the present models indeed hints a QGP onset, 
we explore two distinct effects that might result
in a substantial increase of the transverse pressure: I)
initial state Cronin enhancement and II) heavy resonance formation.

\section*{The Models}
In our studies we use two independent relativistic transport models
that employ hadronic and string degrees of freedom:  UrQMD
\cite{UrQMD1,UrQMD2} and HSD \cite{Geiss,Cass99}. They take into
account the formation and multiple rescattering of hadrons and 
dynamically describe the generation of pressure in the hadronic
expansion phase. This involves also interactions of leading pre-hadrons
that contain a valence quark (antiquark) from a primary 'hard'
collision (cf. Refs.  \cite{Weber02,Cass_Cron03}). Note that, in these models, only hadrons, 
valence quarks and valence diquarks and their interactions are treated explicitly. Gluonic 
degrees of freedom are not treated explicitly, but are implicitly present  in strings.
This simplified treatment is generally accepted to describe proton-proton and proton-nucleus 
interactions. Here we test whether this description is still valid for the more
complicated nucleus-nucleus collisions, where large energy densities can be 
reached over extended volumes.

\section*{Transverse Dynamics in Small and Large Systems}
\begin{figure}[hbt]
%\vspace*{-1.0cm}
\centerline{\psfig{figure=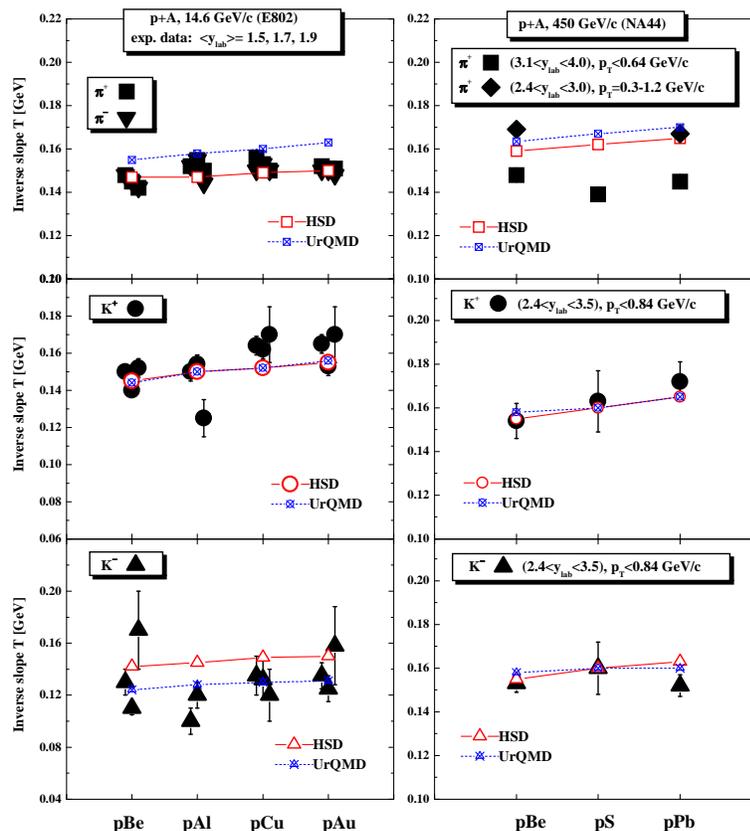,width=10cm}}
%\insertplot{t-pa.eps}
\vspace*{-.5cm}
\caption[]{Inverse slope parameters $T$ for $\pi^\pm, K^+$ and $K^-$
at midrapidity from pA reactions at 14.6 GeV$/c$ (A= Be, Al, Cu, Au)
 -- left part and at 450 GeV$/c$ (A= Be, S, Pb) -- right part, from
HSD (open symbols) and UrQMD 2.0 (closed symbols).
The full symbols in the left part correspond to the midrapidity
data ($\langle y_{lab}\rangle = 1.5, 1.7, 1.9$) from the E802 Collaboration
\protect\cite{E802_pA},
in the right part to the NA44 data \protect\cite{NA44_pA}
at $2.4\le y_{lab}\le 3.5, \ p_T\le 0.84$ GeV$/c$ for $K^+, K^-$ and
at $2.4\le y_{lab}\le 3.0, \ p_T =0.3\div 1.2$ GeV$/c$ (full diamonds)
and $3.1\le y_{lab}\le 4.0, \ p_T\le 0.64$ GeV$/c$ (full squares)
for $\pi^+$. }
\label{Fig_tpa}
\end{figure}

Let us start by ''benchmarking'' the model calculations with pA data.
Fig. \ref{Fig_tpa} shows the results for the inverse 
slope parameters $T$ for various reactions - see figure caption for details.
It can be seen that the models reproduce the transverse slope parameters of different 
particles produced in pA interactions with targets from Be to Pb reasonably well.

\begin{figure}[hbt]
\centerline{\psfig{figure=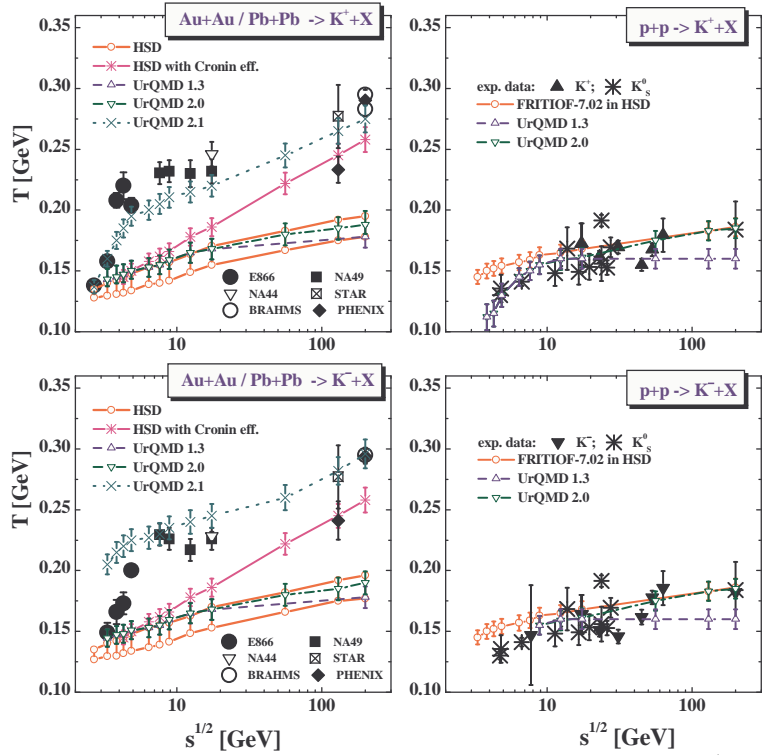,width=10cm}}
%\insertplot{t-s1-cr.eps}
\vspace*{-.5cm}
\caption[]{Comparison of the inverse slope parameters $T$ for $K^+$ and
$K^-$ mesons from central Au+Au (Pb+Pb) collisions (l.h.s.) and pp
reactions (r.h.s.) at midrapidity as a function of the invariant energy
$\sqrt{s}$ from HSD (upper and lower solid lines with open circles),
UrQMD 1.3 (dashed lines with open  triangles),
UrQMD 2.0 (dot-dashed lines with open inverted triangles),
UrQMD 2.1 (dotted lines with crosses) with the data from Refs.
\protect\cite{NA49_T,E866E917,NA44,STAR,BRAHMS,PHENIX} for AA and
\protect\cite{STAR,NA49_CCSi,Gazdz_pp} for pp collisions
The upper and lower solid lines in the left diagrams result from
different limits of the HSD calculations as discussed in the text while
the solid lines with stars correspond to HSD calculations with
the Cronin initial state enhancement.}
\label{Fig_Tcron}
\end{figure}

We continue with nucleus-nucleus collisions, where
Fig. \ref{Fig_Tcron} summarises our results: the
dependence of the inverse slope parameter $T$ 
on $\sqrt{s}$ is shown and compared to (partly preliminary) experimental
data  from 
\protect\cite{NA49_T,E866E917,NA44,STAR,BRAHMS,PHENIX} for central
Au+Au (Pb+Pb) collisions (l.h.s.) and \cite{STAR,NA49_CCSi,Gazdz_pp}
for pp collisions (r.h.s.).  The upper and lower solid lines (with
open circles) on the l.h.s. in Fig. \ref{Fig_Tcron} correspond to
results from HSD calculations, where the upper and lower limits are due
to fitting the slope $T$, an uncertainty in the repulsive
$K^\pm$-pion potential or the possible effect of string overlaps.  The
solid lines with stars correspond to HSD calculations with the
Cronin effect.  The dashed lines with open 
triangles represent slope parameters from UrQMD 1.3, the dot-dashed
lines with open inverted triangles correspond to UrQMD 2.0 results, which
are well within the limits obtained from the different HSD calculations
without the Cronin enhancement.  The dotted lines with crosses show the
UrQMD 2.1 results that incorporate  high mass resonance states up to $m_R\le 3$~GeV.

The slope parameters from pp collisions (r.h.s. in Fig.
\ref{Fig_Tcron}) are seen to increase smoothly with energy both in the
experiment (full symbols) and in the HSD calculations (full lines with
open circles). The UrQMD 1.3 results are shown as open triangles
connected by a solid line and systematically lower than the slopes
from HSD at all energies. When including jet production and
fragmentation via PYTHIA in UrQMD 2.0 (dot-dashed lines with open inverted
triangles) the results become similar to HSD above
$\sqrt{s}$ = 10 GeV demonstrating the importance of jets in pp
reactions at high energy. 

Coming back to the slope parameters of $K^\pm$ mesons for central
Au+Au/Pb+Pb collisions (l.h.s. of Fig.  \ref{Fig_Tcron}) we find that
the Cronin initial state enhancement indeed improves the description of
the data at RHIC energies, however, does not give any sizeable
enhancement at AGS energies.  Here UrQMD 2.1 (dashed lines with
crosses) with the high mass resonance states performs better:
Including high mass resonances one obtains more
reasonable results for $K^+$ mesons, however, fails by 10 to 15\%  for
pions as well as anti-Kaons.  
In this context it is interesting to note that the experimental results on
C+C and Si+Si at 158~AGeV show small slopes \cite{NA49_CCSi} and are therefore in agreement
with the models \cite{MT-prl}.

What is the origin of the rapid
increase of the slopes with energy for central Au+Au
collisions at AGS energies and the constant slope at low SPS energies (the
'step' in the Kaon temperature), which is  missed in presently employed transport approaches?

\section*{Elliptic Flow}
\begin{figure}[hbt]
\vspace*{.2cm}
\centerline{\psfig{figure=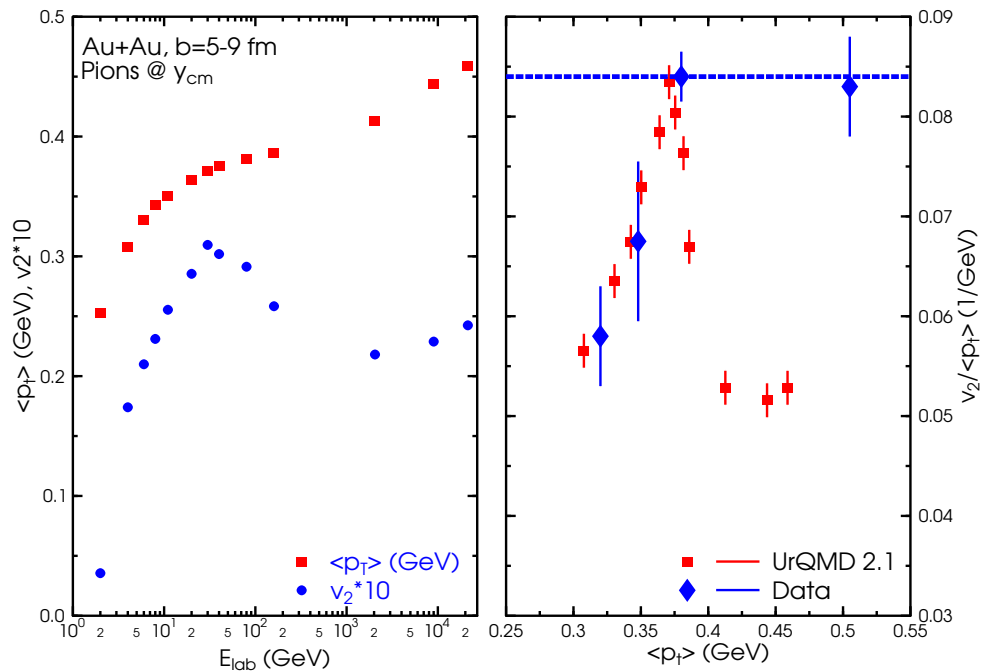,width=13.5cm}}
%\insertplot{tmuurqmd.eps}
\vspace*{-.5cm}
\caption{Left: Calculated excitation functions of mean transverse momentum and elliptic flow of pions 
at midrapidity. Right: Excitation function of $v_2/\langle p_T\rangle$ of Pions in mid-central collisions from top AGS to RHIC energy from UrQMD (squares). 
Negatively charged particle data (diamonds)
for $v_2$ are taken from Ref. \cite{Alt:2003ab} and $\langle p_T\rangle$ from Refs.\cite{Akiba:1996xf,bramm,Appelshauser:1998yb,Adler:2001yq}.}
\label{Fig_mpt}
\end{figure}

To disentangle the effects of high mass hadron states from a possible
phase transition scenario we suggest to study the energy excitation function of the
elliptic flow of pions (or negatively charged hadrons).
The $\langle p_T\rangle$ and $v_2$ excitation functions are 
depicted in Fig. \ref{Fig_mpt} (left). One clearly observes a monotonic rise in the 
mean transverse momentum of the pions with increasing energy.
However, the elliptic flow behaves non-monotonic and shows a distinct maximum around 
30-40 GeV beam energy. This phenomenon can be pronounced more clearly by the scaled elliptic 
flow ($v_2/\langle p_T\rangle$) as shown in Fig. \ref{Fig_mpt} (right). 
It was pointed out \cite{Huovinen:2001wn,snellings} that at high energies the differential
$v_2(p_T)$ of charged hadrons is approximately proportional to $p_T$,
such that the averaged $v_2\propto \langle p_T\rangle$. In fact, 
when divided by the average transverse momentum, the 
scaled elliptic flow in that energy regime becomes nearly constant \cite{snellings,Paech:2003fe}, 
even though the measured $v_2$ increases from SPS energy to RHIC energy.

To emphasise deviations from the natural scaling $v_2\propto \langle p_T\rangle$ , we
plot the excitation function of $v_2/\langle p_T\rangle$ in Fig. \ref{Fig_mpt} (right).
At lower energies there is a systematic increase of $v_2$
relative to $\langle p_T\rangle$ in the model and in the data. 
However, above the SPS energy regime, one clearly observes 
that the data for the scaled elliptic flow is constant and independent of energy while the model yields
first a sharp decrease of the $v_2/\langle p_T\rangle$ which then levels-off at 
roughly half the experimentally observed value.
 
The present non-equilibrium study however, suggest that the initial
increase of the scaled elliptic flow up to SPS energies, might be due to 
viscosity effects (decreasing mean-free-path) in the hadronic gas.  
At higher energies, the predicted elliptic flow
breaks down in the model calculation because of the increasing dominance of string dynamics. 
The measured data however, supports a hydrodynamical behaviour of the matter in the early stage 
with very small mean free paths.
It should be noted that a similar minimum in the elliptic flow was first predicted by hydrodynamics.
In contrast to the scenario discussed here, this decrease of $v_2$ was associated with the softening of
the equation of state in the phase transition region \cite{Kolb:1999it,Kolb:2000sd}.  

Thus, from the lack of initial pressure we conclude that the system (at least at RHIC energies) 
seems to spend a considerable amount of
time in the QGP phase with an equation of state harder than the employed hadron/string 
gas equation of state. 
This argument is well in line with the studies on
elliptic flow at RHIC energies, which is underestimated by $\sim$ 30\%
at midrapidity in HSD  \cite{Brat03} and by a factor of
$\sim 2$ in UrQMD 1.3  \cite{Bleicher_v2}.  It is our opinion that strong
pre-hadronic/partonic interactions might cure this problem.

\section*{Conclusion}
In conclusion, we have found that the inverse slope parameters $T$ for $K^\pm$ mesons
from the HSD and UrQMD 1.3 transport models are practically independent of
system size from pp up to central Pb+Pb collisions and show only a
slight increase with collision energy. The calculated transverse mass
spectra are in reasonable agreement with the experimental results for
pp reactions at all bombarding energies investigated as well as
central collisions of light nuclei (C+C and Si+Si) (cf. Ref.
\cite{MT-prl}). The rapid increase of the inverse slope parameters of
Kaons for central collisions of heavy nuclei (Au+Au or Pb+Pb) found
experimentally in the AGS energy range, however, is not reproduced by
both models in their default version (see Fig.~\ref{Fig_Tcron}).
 
We have discussed  scenarios 
to improve the description of the experimental data.
However, no fully convincing results could be obtained for all observables
and bombarding energies simultaneously.

From comparison to lattice QCD calculations  at
finite temperature and baryon chemical potential $\mu_B$ from Refs.
\cite{Fodor} and \cite{Karsch2} as well as the experimental systematics
in the chemical freeze-out parameters (cf. Fig. \ref{Fig_QCD}), we infer
that
the missing pressure above 30 GeV beam energy might be generated in the early phase of the
collision by non-perturbative partonic interactions.
However, to fully clarify this issue will require a systematic quantitative comparison  
with hydrodynamic models from the lowest AGS energy to the highest RHIC energy.

\noindent
\section*{Acknowledgements}
This work was supported by GSI, DFG and BMBF.
This work used computational resources provided by the
Center for Scientific Computing at Frankfurt (CSC).

\end{document}